\documentclass[pra,aps,twocolumn,floatfix]{revtex4}
\usepackage[dvips,final]{graphicx}
\usepackage{amsmath}

\def\be{\begin{equation}}
\def\ee{\end{equation}}
\def\bea{\begin{eqnarray}}
\def\eea{\end{eqnarray}}

\begin{document}

\title{Optical discrimination between spatial decoherence
and thermalization of a massive object}

\author{C. Henkel$^1$, M. Nest$^2$, P. Domokos$^3$, and
R. Folman$^4$}
\affiliation{%
$^1$Universit\"at Potsdam,
Institut
f\"{u}r Physik,
Am Neuen Palais 10,
14469 Potsdam,
Germany\\
$^2$Universit\"at Potsdam,
Institut f\"ur Chemie,
Karl-Liebknecht-Str. 25, 14476 Potsdam, Germany\\
$^3$Research Institute for Solid State Physics and Optics, Budapest,
Hungary\\
$^4$Department of Physics and Ilse Katz Center for Meso- and Nanoscale
Science and Technology, Ben Gurion University of the Negev, P.O.Box
653, Beer Sheva 84105, Israel}

\date{02 June 2004}


\begin{abstract}
We propose an optical ring interferometer to observe
environment-induced spatial decoherence of massive objects. The
object is held in a harmonic trap and scatters light between
degenerate modes of a ring cavity. The output signal of the
interferometer permits to monitor the spatial width of the
object's wave function. It shows oscillations that arise from
coherences between energy eigenstates and that reveal the
difference between pure spatial decoherence and that coinciding
with energy transfer and heating. Our method is designed to work
with a wide variety of masses, ranging from the atomic scale to
nano-fabricated structures. We give a thorough discussion of its
experimental feasibility.
\end{abstract}

\maketitle



\section{Introduction}

The Schr\"odinger cat is a well known example of the difficulty in
clearly defining the border between the classical and quantum
worlds. In this example, quantum theory allows for a macroscopic
superposition of a dead and a live cat to exist while we have never
been able to observe such a macroscopic superposition in
nature. Indeed, this enigma has been the source of a century long
debate.

To the best of our knowledge, only few attempts have been made so far
to artificially create macroscopic superposition states.  One attempt
dealt with a superposition of two states of a multi-photon cavity
field \cite{brune}. A second dealt with a superposition of a magnetic
flux direction, formed by two macroscopic counter propagating electron
currents \cite{flux}.  Spatially separated superpositions of a trapped
ion have been prepared and probed~\cite{monroe}, and Bose Einstein
condensates have been examined \cite{bec}.  Work on handedness of
chiral molecules may also be considered relevant to this topic
\cite{vager}.  Microscopic, mechanical oscillators have been
extensively discussed as well~\cite{folman, knight}, and experiments
are now entering the regime where quantum effects become
observable~\cite{mems}.

In this paper, we discuss a general oscillating massive object and
its decoherence in the position basis, the basis which stands at
the base of our classical perception.  This spatial decoherence,
also called ``localization'', is predicted by numerous models for
environment-induced decoherence that are put forward to explain
the appearance of classical reality from an underlying quantum
world \cite{dec}: objects become localized in position due to
their interaction with the environment.  Of special interest is
``pure'' decoherence where localization can happen even without
the transfer of energy, e.g., in a double-well potential.  In
order to reproduce the absence of macroscopic superpositions, most
decoherence models use the mass of the decohering object as a
central parameter along with parameters such as time and spatial
splitting of the superposition. Indeed, recent seminal experiments
on matter wave diffraction have tried for the first time to
explore a region of mass values beyond the usually experimentally
feasible masses of elementary particles \cite{anton}. The
visibility loss in the diffraction pattern is then a measurable
signal of spatial decoherence.

Interference experiments with freely propagating objects, however,
become increasingly hard to perform with larger masses for two
main reasons: first, the de Broglie wave length becomes smaller
and the required diffraction gratings become more difficult to
fabricate. Second, the spatial superposition created by
diffraction is increasingly sensitive to decoherence because it is
more difficult to isolate the freely propagating system from its
environment. This renders a controlled experiment with large
masses extremely difficult.

In this paper we propose an interference experiment in which no
grating and no free propagation are needed. In fact, we avoid all
together the need to create a well separated spatial
superposition, and hence the way should be open to perform
controlled decoherence experiments with large masses. Our
experiment is based on an optical interferometer to probe the
state of an oscillating mass, e.g.\ a nano bead or a mechanical
oscillator. Contrary to previous work concerning oscillating
mirrors~\cite{knight,optomech}, a symmetric ring interferometer is
used, and the oscillator is not required to have a high
reflectivity. This feature is advantageous when very light (thin)
nano objects are investigated, as high transmittance does
not pose a problem. We show that this setup can distinguish
between different models of decoherence dynamics so that
information about this subtle process can be obtained
experimentally.

The next section describes the experimental setup.  The
theoretical model is presented in Sect.\ III, and solved
approximately in Sect.\ IV.  In Sect.\ V the experimental
requirements are evaluated for specific examples. General
conclusions are put in Sect.\ VI.

\section{Experimental setup}

\begin{figure}[hbt]
\begin{center}
\includegraphics[width=8cm]{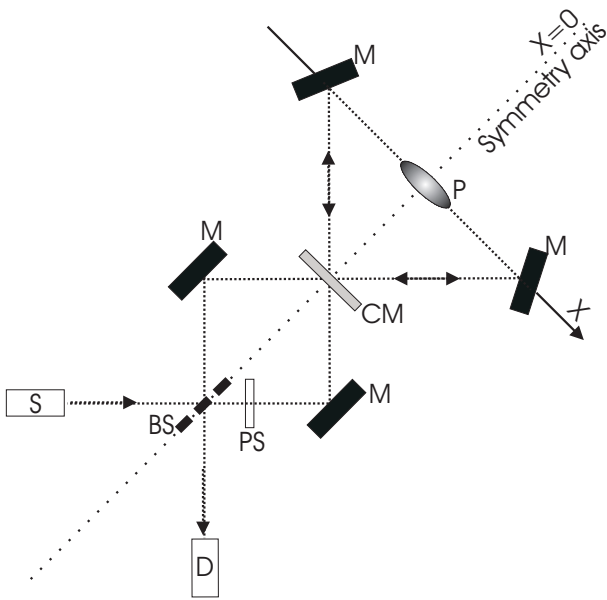}
 \caption{Setup of the experiment. An atom or a more massive oscillating
object, e.g.\ a
   nanoparticle, is held in a harmonic
   potential (P). An incoming photon is split and ``hits'' the object
   from both sides. The source (S), the beam splitter (BS) and the
   phase shifter (PS) form a preparation system which ensures that the
   photon mode is symmetric with respect to the symmetry axis. (PS
   compensates the phase difference between the reflected and
   transmitted wave at the beam splitter.) The same system acts as a
   detection system whereby the antisymmetric photon mode is sent to
   detector D. The potential is symmetric about the symmetry axis as
   well. If, for example, the object is initially prepared in a state
   of well-defined parity (e.g.\ the ground state of the potential),
   its final state will remain a parity eigenstate unless decoherence
   breaks the initial symmetry of the photon+object system.}
 \label{setup}
\end{center}
\end{figure}

The experimental setup we propose is sketched in Fig.~\ref{setup}:
an object of mass $m$ is confined in a potential (P) which can for
all practical purposes be arranged in such a way that only the
motion along one direction is relevant, the $x$-axis, say. We
assume in this paper a harmonic confinement,
$V(x)=m\Omega^2x^2/2$. The object is held at the center of a ring
interferometer, formed by massive mirrors (M) and an in-out
coupling mirror (CM). The symmetry axis of the harmonic potential
coincides with the plane of the beam splitter (BS), and hence with
the symmetry axis of the whole setup. The BS and phase shifter
(PS) prepare a symmetric photon mode, which excites a superposition
(with zero phase shift) of right and left circulating travelling
waves in the
ring cavity. The BS and PS also act as a measurement apparatus for the
outgoing modes, sending all anti-symmetric modes into the detector
D. This is so because anti-symmetric modes of the cavity, are
constructed from right and left circulating
beams with a $\pi$ phase shift. These are coupled out of the
cavity through the CM in two separate directions. The PS 
reduces their phase difference to $\pi/2$ as in a normal
Mach-Zehnder apparatus, whereby the BS determines
constructive interference in the direction of D.

The experimental procedure we have in mind is the following: at
$t<0$ the object is prepared in some equilibrium state at a given
temperature. At $t=0$ the preparation stops, and the environment,
be it a thermal bath or some tailored environment, becomes
dominant in the dynamics of the system. At $t=t_p$, a probe light
pulse is sent into the ring interferometer and interacts with the
object. The consequent measurement of the probe pulse which exits
the interferometer after the interaction, determines if the even
symmetry of the initial photon has been altered. In principle also
the energy change of the outgoing photon may be measured, but in
this paper we make no use of this option.

We note that the operation of the ring interferometer is not
qualitatively affected when working with a ``transparent'' object like
an atom or a weakly scattering nanoparticle. For the ease of
demonstration we first describe a single atom. The potential
could be provided in this case by a magnetic trap \cite{folman1}.
Following that, we extend the treatment to a massive 
nanoparticle to show how the experimental sensitivity scales 
with mass and temperature.

To conclude, the scheme provides an experimental measure of
decoherence which may be used for probing large mass objects. It is
interesting to note that future technologies may enable confinement of
massive objects also in potentials other than harmonic, in which case
decoherence can lead to a clearer signature of spatial localization
without energy transfer (``pure'' decoherence). However, in this work,
we show that already for the most feasible of large mass potentials
(the harmonic well), the ring interferometer is able to distinguish
between different decoherence scenarios, and hence we truly present a
realizable scheme for the probing of decoherence with large masses.

\section{Model}

We use standard techniques of open system quantum dynamics to model
the experiment sketched in the previous section. The initial
state of the system (atom or nanoparticle) is described by
a density matrix $\rho_A$, which represents thermal
equilibrium in the harmonic trapping potential at temperature
$T_0$. Up to some probing time $t = t_p$, the system evolves
according to a Liouville-von Neumann equation
\begin{equation}
\dot{\rho}_A = -\frac{i}{\hbar} [H_S,\rho_A] + {\cal L}[\rho_A]
,
\label{eq:Liouville-Neumann}
\end{equation}
with the harmonic oscillator Hamiltonian
\begin{equation}
H_S = \frac{p^2}{2m}  + \frac{m\Omega^2}{2}x^2
\label{eq:HS-harmonic-oscillator}
\end{equation}
and a dissipative functional ${\cal L}$ that describes
the influence of the environment on the system. Its expression
is detailed below.

Around time $t_{p}$, a light pulse is injected into the cavity to
probe the state of the oscillator.  The pulse sent into the cavity has
a center frequency close to a cavity resonance and a narrow bandwidth
compared to the free spectral range (FSR, which will be denoted by
$\nu$). We assume that the oscillation frequency $\Omega$ is much
smaller than $\nu$ so that the interaction of the system with the
field can only couple degenerate cavity modes.  For simplicity, we
also neglect the light scattered by the object into higher transverse
cavity modes.  The field in the cavity can then be described by only
two degenerate modes with even and odd symmetry. As shown in
Appendix~A, the interaction Hamiltonian is given by
\begin{multline}
H_{int} =  \hbar g \left(
\left( a_e^\dagger a_e - a_o^\dagger a_o \right) \cos (2kx)
+ \right. \\
 \left. {}
+\left( a_e^\dagger a_o +a_o^\dagger a_e \right) \sin (2kx) \right)
\label{eq:Hint}
\end{multline}
where $a_e$ and $a_o$ are the boson operators for the even and odd
cavity modes, respectively, $k = 2\pi/\lambda$ is the cavity wave
number, and $g$ is the coupling strength. The excitation of the
even mode is governed by the term
\begin{equation}
  \label{eq:pump}
  H_{pump} = -i \hbar \left( a_e^\dagger \eta(t) - \eta^\dagger(t) a_e \right)\; ,
\end{equation}
where the pump amplitude $\eta$ is an operator whose state allows to
describe any photon statistics of the incoming pulse.  The leakage of
photons out of the cavity is determined by the transmissivity of the
coupling mirror (CM). This process is accounted for by the Liouville
functional
\begin{equation}
{\cal L}^{cav} [\rho_{AF} ] = 2\kappa \sum_{i=e,o}\left( a_i \rho_{AF}
a_i^\dagger - \frac{1}{2} [a_i^\dagger a_i, \rho_{AF} ]_+ \right) \; ,
\end{equation}
where $\kappa$ is the finite cavity linewidth (HWHM), $[\cdot,
\cdot]_{+}$ denotes the anti-commutator, and $\rho_{AF}$ is the joint
density operator of the system and the cavity field modes.  We assume
for simplicity that all photons leaking out are actually detected with
unit efficiency. Thus, the photon rate at the ``odd'' detector (D) is
given by $2\kappa$ times the number of odd photons in the cavity.

As a signal, we will consider the average photon number detected
within in the time interval $[t_p,t_f]$, given by
\begin{equation}
N_o = 2 \kappa \int_{t_p}^{t_f} {\rm Tr}\left( a_o^\dag a_o^{\phantom\dag}
\rho_{AF}(t) \right) dt
\;.
\label{eq:odd-photon-number-0}
\end{equation}
The upper bound of the integration, $t_f$, is chosen such that the
detected photon rate is negligibly small by this time.  The pulse
length, and hence the detection window, is kept short compared to the
decoherence timescale of the system.

We focus in this paper on two different dissipative functionals that
describe opposite extremes of dissipation and decoherence. We require the
time-evolution to be a completely positive semi-group, i.e. of Lindblad
type~\cite{alicki}. The functional
\begin{equation}
{\cal L}^{dec} [\rho_A ] = - \frac{D}{\hbar^2} [x,[x,\rho_A]]
\label{eq:dephasing}
\end{equation}
describes the limit of pure spatial decoherence ($D$ is the classical
momentum diffusion coefficient). It corresponds to the limit of
weak friction and high-temperature environment in a Caldeira-Leggett
model (see, e.g., \cite{Tannor97}). We obtain a pure thermalization
Liouvillian by adding the requirement of detailed balance to the complete
positivity. In terms of the creation and annihilation operators of our
harmonic system Hamiltonian~(\ref{eq:HS-harmonic-oscillator}),
\begin{eqnarray}
{\cal L}^{therm} [\rho_A ] & = & \Gamma_{\downarrow} \left( b \rho_A
b^\dagger - \frac{1}{2} [b^\dagger b, \rho_A ]_+ \right)
\label{eq:thermalization}\\
\nonumber {} & {} & + \Gamma_{\uparrow} \left( b^\dagger \rho_A b -
\frac{1}{2} [b b^\dagger, \rho_A ]_+ \right)
.
\end{eqnarray}
The stationary state of this functional
is a canonical ensemble, with temperature
$k_B T_0 = \hbar \Omega / \ln( \Gamma_{\downarrow}
/ \Gamma_{\uparrow} )$ \cite{dec,alicki}.
In both cases,
the description in terms of a master equation is a reasonable choice
if the coupling to the environment is weak.

Pure spatial decoherence can localize a system without the transfer of
energy, for example in a deep double-well potential or through
recoil-free scattering of probe particles. In quantum Brownian motion,
this corresponds to the limit where the environment correlation time
is so short that the rotating wave approximation with respect to the
system's oscillation frequency cannot be made and terms like $b^2$ and
$b^{\dag2}$ are retained in the interaction with the
environment~\cite{Tannor97}.  These terms drive the system from the
initial ground state $\rho_{A}( 0 ) = | 0 \rangle \langle 0 |$ to a
squeezed state. Therefore, after some time $t_p>0$, its density matrix
in the energy basis will contain higher excited states such that
off-diagonal elements become populated. Thermalization
~(\ref{eq:thermalization}), on the other hand, simply redistributes
the weights of the diagonal elements $\langle n | \rho_A | n \rangle$.
The coherences of the energy eigenstates produce a breathing motion of
the spatial density which can be detected by the probe pulse as $t_p$
is varied. This signal then distinguishes spatial decoherence from
thermalization, and allows us to probe pure decoherence for a massive
body.

\section{Analytical solution}

In this section, we work out the system density operator under the
action of the decoherence models of the previous section, and analyze
how it leaves a characteristic trace in the cavity mode operators.

\subsection{Approximations}
\label{s:approximations}


We summarize first the approximations we make to arrive at
an analytical solution.

(i) The most significant difference of our approach compared to
related work on mobile mirrors is the ``sudden approximation'': we
assume that the duration $\tau$ of the pulse is short compared to the
system oscillation period. Since the damping rate $\kappa$ also
determines the actual pulse length inside the cavity, we require
\begin{equation}
    \Omega \ll \min( 1/\tau, \kappa).
    \label{eq:condition-sudden}
\end{equation}
In this limit, the system's motion is ``frozen'' while the pulse is
applied, and the Heisenberg equations for the photon mode operators
can be solved without taking into account the dynamics of the system
operators.

(ii) At the same time, the pulse must be sufficiently long in order to
restrict the cavity dynamics to the two degenerate modes mentioned
previously.  This is valid when the inverse pulse length is small
compared to the cavity free spectral range $\nu$.  Combining this with
the ``bad cavity limit'' assumed below ($\kappa \tau \gg 1$), but
excluding a too small cavity finesse, we have
\begin{equation}
    1/ \tau \ll \kappa < \nu.
    \label{eq:condition-two-mode}
\end{equation}

(iii) Decoherence can be described by a dissipative functional as in
Eq.~(\ref{eq:Liouville-Neumann}) if the system is weakly coupled to
its environment and decoherence is happening slowly on the timescale
set by the oscillation period $2\pi/\Omega$. This requires the
inequality
\begin{equation}
    \Gamma_{\rm th} \ll \Omega
    \label{eq:condition-master-eqn-1}
\end{equation}
for the thermalization model~(\ref{eq:thermalization}).  For the pure
spatial decoherence model~(\ref{eq:dephasing}), we require that it
takes more than one oscillation to increase the average system energy
by one quantum $\hbar\Omega$.  This leads to
\begin{equation}
    \Gamma \ll \Omega,
    \label{eq:condition-master-eqn-2}
\end{equation}
where the rate
\begin{eqnarray}
    &&\Gamma \equiv \frac{ D }{\hbar \Omega m }
\label{eq:def-Gamma-dec}
\label{eq:def-decoherence-rate}
\end{eqnarray}
gives the depletion of the system ground state, see e.g.\
Folman et al.\ in~\cite{folman1}.
Given Eq.\ (\ref{eq:condition-master-eqn-2}), depletion happens
slowly on the scale of the oscillation period.
In both cases, the oscillator has a large quality factor.

\subsection{Decoherence}

We now show that the dissipative evolution can be integrated in terms
of the system covariances in position and momentum. For the
equilibrium initial state we consider here, the mean values $\langle x
\rangle$, $\langle p \rangle$ vanish at all times by symmetry. Using
the spatial decoherence functional~(\ref{eq:dephasing}), we get from
the Liouville-von Neumann equation~(\ref{eq:Liouville-Neumann})
\begin{subequations}
 \label{eq:ddt_cov}
\begin{align}
\frac{ {\rm d} }{ {\rm d}t }
\langle x^2 \rangle &=
\frac{1}{m}
\langle p x + x p \rangle
\label{eq:ddtx2}
\\
\frac{ {\rm d} }{ {\rm d}t }
\langle p^2 \rangle &=
- m \Omega^2
\langle p x + x p \rangle
+ 2 D
\label{eq:ddtp2}
\\
\frac{ {\rm d} }{ {\rm d}t }
\langle x p + p x \rangle &=
\frac{ 2 }{ m } \langle p^2 \rangle
- 2 m \Omega^2 \langle x^2 \rangle
.
\label{eq:ddtxp}
\end{align}
\end{subequations}
Characteristic for spatial decoherence is that only the momentum width
is increased by the diffusion coefficient $D$.  This ``squeezes'' the
system state in the phase plane, while the dynamics in the harmonic
potential subsequently leads to a rotation.  The coupled
equations~(\ref{eq:ddtx2}--\ref{eq:ddtxp}) can be solved, with the
result (see, e.g., \cite{Tannor97})
\begin{subequations}
\begin{eqnarray}
\langle x^2 \rangle_t &=&
\langle x^2 \rangle_0 +
\frac{ \hbar \Gamma }{ 2 m \Omega^2 }
\left( 2 \Omega t - \sin 2 \Omega t \right)
\label{eq:solution-x2}
\label{eq:result-sudden-2}
\\
\langle p^2 \rangle_t &=&
\langle p^2 \rangle_0
+
\frac{ \hbar m \Gamma }{ 2 }
\left( 2 \Omega t + \sin 2 \Omega t \right)
\\
\langle x p + p x \rangle_t &=&
\frac{ \hbar \Gamma }{ \Omega }
\left( 1 - \cos 2 \Omega t \right)
,
\end{eqnarray}
\end{subequations}
where the decoherence rate $\Gamma$ has been defined
in~(\ref{eq:def-decoherence-rate}).  We have used that equipartition
holds in the initial state, $\langle p^2\rangle_0 / m = m \Omega^2
\langle x^2 \rangle_0$, which is obviously true for an initial thermal
state. Note that the decoherence Liouvillian does not describe a
stationary solution in the limit $t \to \infty$, this is because we
neglected friction.

For the thermalization model~(\ref{eq:thermalization}),
the variances can be computed similarly, and we find
\begin{subequations}
\begin{eqnarray}
\langle x^2 \rangle_t &=& \langle x^2 \rangle_0 + \frac{ E(T_e) }{ m \Omega^2 }
\left( 1 -  e^{ - 2 \Gamma_{\rm th} t}  \right)
\label{eq:solution-x2-thermalization}
\label{eq:result-sudden-3}
\\
\langle p^2 \rangle_t &=& \langle p^2 \rangle_0 + E(T_e) m
\left( 1 -  e^{ -2 \Gamma_{\rm th} t}  \right) \\
\langle x p + p x \rangle_t &\equiv& 0 \\
\Gamma_{\rm th} &=&  \frac{ \Gamma_{\downarrow} - \Gamma_{\uparrow} }{ 2 }
,
\end{eqnarray}
\end{subequations}
where $E(T_e)=\frac{1}{2} \hbar\Omega \coth( \hbar\Omega / 2 k_B T_e)$ is
the average oscillator energy at equilibrium with the environment
(temperature $T_e$), and where the rate $\Gamma_{\rm th}$ characterizes
both the approach towards thermal equilibrium and the damping of the
system's average position and momentum.


The key benefit of this formulation in terms of covariances is that it
provides an exact solution for the system density
operator~\cite{Haake85,Ford01b}. The reasons for this are the thermal
initial state we consider here and the Liouville
functionals~(\ref{eq:dephasing}, \ref{eq:thermalization}) that are
bilinear in $x, \,p$.  We shall work with the Wigner representation
$W(x, p, t)$ of the density operator that has properties similar to a
classical phase space distribution \cite{VanKampen}. For example,
expectation values of symmetrized system operators $S(\hat x,\hat p)$
are computed according to
\begin{equation}
\langle S(\hat x,\hat p) \rangle_t
=
\int\!\frac{ dx\,dp }{ 2\pi\hbar }
S(x, p)
W(x, p; t )
.
\label{eq:expectation-value-in-Wigner}
\end{equation}
Given the covariances, we find the Wigner function
\begin{subequations}
\begin{eqnarray}
W(x, p; t ) &=& {\cal N}(t) \exp[ - G(x, p; t ) ]
\\
2 G(x, p; t ) &=&
\frac{ x^2 }{ \langle x^2 \rangle_t }
+
\frac{ 1 }{ \langle p^2 \rangle_t }
\left( p - \frac{ C_t x }{ A_t } \right)^2
\label{eq:exp-Wigner}
\\
\frac{ C_t }{ A_t } &=&
\frac{ \langle x p + p x \rangle_t }{
2 \langle x^2 \rangle_t }
\\
\frac{ 1 }{ {\cal N}(t)^2 } &=&
\frac{ \langle x^2 \rangle_t \langle p^2 \rangle_t
-
{\textstyle \frac14} \langle x p + p x \rangle_t^2 }{
\hbar^2 }
.
\end{eqnarray}
\end{subequations}
Note that for the normalization factor, one finds
${\cal N}(t) \le 2$ because of
the uncertainty relations.

\subsection{Short probe pulse}

In a frame rotating at the cavity resonance frequency, the Heisenberg
equations for the photon operators are
\begin{subequations}
\begin{align}
\dot a_e &= -{\rm i} g \left(
\cos(2kx) a_e + \sin(2kx) a_o \right) - \kappa
a_e + \eta(t) + \xi_e
\\
\dot a_o &= {\rm i} g \left(
\cos(2kx) a_o - \sin(2kx) a_e \right)- \kappa
a_o + \xi_o
\end{align}
\end{subequations}
where $\xi_{e,o}$ are quantum noise operators that can be neglected as
long as we calculate normally ordered quantities, such as the intensity.
In the sudden approximation, we assume that the system position
operator $x$ does not change during the pulse duration. It can then be
treated as a constant that commutes with the photon operators.

The initial condition just before time $t_p$ is vacuum for both cavity
modes.  The evolution of the odd mode is then given by (neglecting
terms with vanishing expectation value)
\begin{multline}
a_o(t_p+t) =
- i \sin(2 k x) \\ \times \frac{g - g e^{-\kappa t }
  \cos{g t} - \kappa  e^{-\kappa t } \sin{g t}}{g^2+\kappa^2} \,
  \eta
\label{eq:result-a-odd}
\end{multline}
for $ 0 \le t \le \tau$, where $\tau$ is the pulse length, and the
pulse shape was taken as a mesa-function: $\eta(t_p + t) = \eta$ for
$0< t< \tau$, otherwise it vanishes. With this choice,
$\eta=\sqrt{\kappa/(2\tau)}\, a_{in}$, where $a_{in}$ is the boson
operator of the probe pulse incident on the cavity.

The average number of odd photons $N_o$ at the detector during the
time window $[t_p,t_f]$ can be calculated by integrating $2\kappa
a^\dag_o(t)a_o(t)$, see Eq.~(\ref{eq:odd-photon-number-0}).  As
mentioned in Sec.~\ref{s:approximations}, we consider here the bad
cavity limit where $\kappa \tau \gg 1$. In this limit, we may set $t_f
= t_p+\tau$, and the integration yields the simple result
\begin{equation}
N_o = R \langle a^\dag_{in}a_{in}\, \sin^2(2 k x) \rangle
\label{eq:Nodd-result}
\end{equation}
with
\begin{equation}
    R = \frac{g^2\kappa^2}{(g^2+\kappa^2)^2}
    \; .
    \label{eq:def-R}
\end{equation}
The average over the field and system operators in
Eq.~(\ref{eq:Nodd-result}) can be computed independently because we
have assumed that the system+field density operator factorizes at $t =
t_p$.  The signal is thus proportional to the mean photon number of
the probe pulse, $\langle a^\dag_{in}a_{in} \rangle = N_{in}$.  We
discuss the signal fluctuations in Sec.~\ref{s:signal-fluctuations}
below.  For the average over the system, the Wigner
function~(\ref{eq:exp-Wigner}) gives after one elementary integration
\begin{eqnarray}
\langle \sin^2(2kx)\rangle_{t_{p}}
&=&
\int\!\frac{ dx\,dp }{ 2\pi\hbar }
\sin^2(2kx)
W(x, p; t_{p} )
\nonumber
\\
&=&
\frac{ 1 - \exp\left[ - 8 k^2
\langle x^2 \rangle_{t_{p}} \right] }{ 2 }
\;,
\label{eq:sin2-and-k2}
\label{eq:result-sudden-1}
\end{eqnarray}
where the variance $\langle x^2 \rangle_{t}$ has been calculated in
Eqs.\ (\ref{eq:solution-x2}) and (\ref{eq:solution-x2-thermalization}).

We observe that the solution~(\ref{eq:result-a-odd}) illustrates how our
detection scheme conserves parity.  Consider an incident pulse in a single
photon state, $|\Psi (t_p) \rangle = a_{in}^\dag | 0 \rangle$, and
assume that the odd detector clicks.  This updates the system state to
\begin{equation}
    \mbox{``odd click'':}\quad
    \rho_{A} \mapsto {\cal K} \sin(2kx) \rho_{A} \sin(2kx) \; ,
\end{equation}
where ${\cal K}$ is the normalization.  If the system has been in a
state $\rho_{A}$ with definite parity, this state has the opposite
one.  Similarly, a click in the even mode detector updates the system
state to a state with the same parity.  In both cases, the
``collapse'' of the wave function does not lead to an a-symmetric
state with less well defined parity.  This ensures that symmetry of
the system state can only be changed by decoherence.

The factor $R$ in Eq.\ (\ref{eq:def-R}) gives basically the probability
that an incident photon actually interacts with the system. This can
be seen from the number of even photons that is given by
\begin{equation}
    N_{e} = \bar R N_{in}
    + R N_{in} \langle \cos^2(2 k x) \rangle
    \label{eq:even-photons}
\end{equation}
where $\bar R = {\kappa^4 }/{(g^2 + \kappa^2)^2}$.  The first term
is independent of the system position and gives the photons that
did not interact with the system. This number reduces to $N_{in}$
for a vanishing coupling, $g \to 0$. The second term has a similar
structure as Eq.\ (\ref{eq:Nodd-result}), with the difference that
here only the system operator with even parity occurs. It adds up
with $N_{o}$ to $R N_{in}$, similar to the two output ports of an
interferometer. The fraction of `useful' photons is thus $R / (R +
\bar R) = g^2/(g^2 + \kappa^2)$.

For typical experimental conditions that we discuss in
Sec.\ref{s:parameters}, the probe wavelength is large compared to the
width of the system position distribution (``Lamb-Dicke limit'').  We
can then expand the exponential in Eq.\ (\ref{eq:result-sudden-1}) to
get
\begin{eqnarray}
    N_{o} \approx 4 R N_{in} k^2 \langle x^2\rangle_{t_{p}}
    \;,
    \label{eq:sin2-LambDicke}
    \label{eq:signal-2}
\end{eqnarray}
so that a larger signal is obtained with a shorter probe wavelength.
This scaling breaks down, however, in the extreme case of $\lambda$
being comparable or shorter than the position width (``anti Lamb-Dicke
limit''). This may also occur in the long-time limit, after heating
has significantly broadened the position distribution. The signal then
no longer increases linearly with $\langle x^2\rangle_{t_{p}}$. The
exponential in Eq.~(\ref{eq:result-sudden-1}) vanishes, and the signal
saturates at $N_o = R N_{in}/2$ regardless of the value of the
$\langle x^2\rangle_{t_{p}}$. In this limit, the probe pulse is no
longer able to extract information about the system.

We note that the same result $N_o = R N_{in}/2$ may be arrived at on
much shorter time scales (even for $t_p = 0$) when the initially
prepared system is already in the anti Lamb-Dicke limit. The system
then heats as a result of photon scattering (``back action'').  The
exponential $\exp( - 8 k^2 \langle x^2 \rangle_{0} )$ in
Eq.~(\ref{eq:result-sudden-1}) is in fact the Debye-Waller factor for
this scattering process.  It gives the probability of the system still
occupying the ground state after the photon pulse has impinged on it.
In the anti Lamb-Dicke limit, the Debye-Waller factor is zero, and the
system makes with probability unity a transition to a vibrationally
excited state.  Since the detection of an odd photon is directly
correlated, due to symmetry conservation, to a transition from the
even ground state to an odd state, it occurs for one half of all
`useful' photons $R N_{in}$ that actually interacted with the system.

One may also understand this result in the position basis. In the anti
Lamb-Dicke regime, the system position is smeared out over many probe
wavelengths so that the phase of the backscattered light varies
randomly from $0$ to $2\pi$. On average, one half of those photons
that interacted with the system are sent into the odd detector.

\subsection{Signal fluctuations}
\label{s:signal-fluctuations}

For the discussion of the signal to noise ratio in Sec.~\ref{s:s/n},
we also need the fluctuations of the odd detector
signal~(\ref{eq:Nodd-result}).

The variance $\Delta N_{o}^2$ can be computed from the factorized state
$\rho_{AF}( t_{p} )$ as well.  For a probe pulse in a coherent state,
we find
\begin{multline}
    \Delta N_{o}^2 = R^2 \Big(
    N_{in}(N_{in} + 1) \langle \sin^4( 2 k x ) \rangle_{t_{p}}\\
    - N_{in}^2 \langle \sin^2( 2 k x ) \rangle_{t_{p}}^2 \Big)
    \label{eq:Nodd-variance}
\end{multline}
and
\begin{align}
    \langle \sin^4( 2 k x ) \rangle_{t_{p}} &=
    \frac{1}{8}
    \left( 3 - 4\,e^{ - 8 k^2 \langle x^2 \rangle_{t_{p}} } +
    e^{ - 32 k^2 \langle x^2 \rangle_{t_{p}} } \right)\nonumber \\
    & \approx 3 (4 k^2 \langle x^2 \rangle_{t_{p}})^2
    \;.
    \label{eq:sin4-average}
\end{align}
In the last step, we made the long-wavelength expansion as in
Eq.\ (\ref{eq:sin2-LambDicke}).

If the system were located at a fixed position, the terms proportional
to $N_{in}^2$ in Eq.\ (\ref{eq:Nodd-variance}) would cancel, leading to
number fluctuations limited by shot noise. This is not the case here,
however, because of the finite position uncertainty. For the Gaussian
distribution at hand, the fluctuations of $x^2$ are of the same order
and even somewhat larger than their mean value $\langle x^2 \rangle$
itself.  In the long-wavelength limit, Eq.\ (\ref{eq:Nodd-variance})
becomes
\begin{equation}
    \Delta N_{o}^2 = R^2 (4 k^2 \langle x^2 \rangle_{t_{p}})^2
    \left( 2 N_{in}^2
    +
    3 N_{in} \right)
    \;.
    \label{eq:Nodd-variance-2}
\end{equation}
For completeness, we also give the opposite limit of a short
wavelength, although it is experimentally more challenging.  Recall
that the average signal $N_{o} \approx R N_{in} / 2$ arises because
the phase $\phi = 2 k x$ is uniformly distributed between $0$ and
$2\pi$, and the fraction of photons in the odd detector proportional
to $\langle \sin^2\phi \rangle = \frac12$.  The variance then becomes
$\Delta N_{o}^2 = R^2 ( \frac18 N_{in}^2 + \frac38 N_{in} )$, where
the second term is due to shot noise and the first is due to the
variance $\langle \sin^4\phi \rangle - \langle\sin^2\phi\rangle^2 =
\frac18$. This estimate agrees with the anti Lamb-Dicke limit of the
general Eq.\ (\ref{eq:Nodd-variance}).

\section{Discussion}

We show in this section that experimental conditions should exist
where it is possible to spot the difference between spatial
decoherence and thermalization by measuring the photon signals outside
the cavity.

\subsection{Example I: Single atom}
\label{s:parameters}

\begin{table}
    \begin{tabular}{|l|c|}
        \hline
        vibration frequency $\Omega / 2\pi$ & 50\,{\rm kHz}
        \\
        pulse length $\tau$ & 20\,{\rm ns}
        \\
        cavity line width $\kappa / 2\pi$ & 50\,{\rm MHz}
        \\
        free spectral range $\nu$ & 1\,{\rm GHz}
        \\
        wavelength $\lambda$ & 795\,{\rm nm}
        \\
        Lamb-Dicke parameter $(k\sigma_{0})^2$ & 0.073
        \\
        coupling strength $g/2\pi$ & 10\,{\rm kHz}
        \\
        decoherence rate $\Gamma/2\pi$ & $1\,{\rm kHz}$
        \\
        \hline
        \end{tabular}
        \caption[]{Realistic experimental parameters for a Rb atom
        fulfilling the validity conditions of the calculation, and
        yielding a signal to noise ratio larger than 3 in a spectral
        analysis (as discussed around Eq.(\ref{eq:S/N-2})).}
        \label{t:parameters}
\end{table}

The system we consider in the first example is a single rubidium atom
trapped in a tightly confining magnetic trap.  The chosen parameters
are listed in Table~\ref{t:parameters}. The oscillation frequency is
similar to those achieved with magnetic traps on atom chips
\cite{folman1}. For such trapping frequencies, ground state cooling
leads to a spatial width $\langle x^2 \rangle_0^{1/2}$ of the order of
$\sigma_{0} \equiv (\hbar / 2 m \Omega)^{1/2} \approx 34\,{\rm nm}$.
Trapped ions cooled to the ground state with sideband cooling are also
a good system for this purpose.  We note that a cold thermal state
would be sufficient as well. For the sake of this example, we choose a
relatively long wavelength close to the D1 rubidium line.  This gives
a resonantly enhanced AC polarizability, while absorption and
spontaneous emission can still be minimized with a detuning of several
linewidths.  The corresponding coupling strength $g$ is estimated in
the Appendix.  Pulse length, cavity linewidth and free spectral range
are chosen to comply with the approximations defined by
Eqs.~(\ref{eq:condition-sudden}) and (\ref{eq:condition-two-mode}).

The last quantity, the decoherence rate $\Gamma$, depends on the
specifics of the system.  For an atom in a miniaturized
electromagnetic trap on an atom chip, estimates for heating due to
magnetic field fluctuations are in the range of $\Gamma \sim 1\,{\rm
  s}^{-1}$ \cite{folman1}. Other environmental perturbations can be
added at will to enhance this rate in a controlled way. In fact, a
tunable decoherence source is suitable for the unambiguous,
experimental discrimination between decoherence and
thermalization-induced dynamics.

\begin{figure}
\begin{center}
\includegraphics[width=8cm]{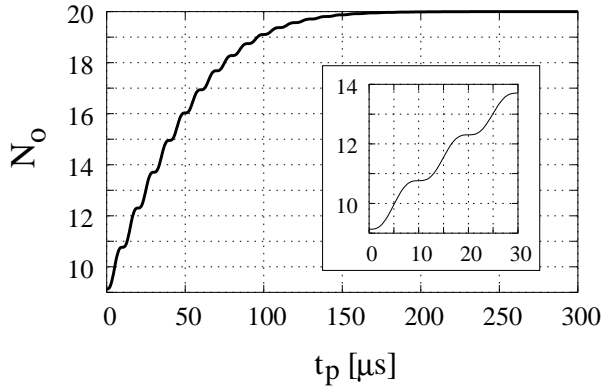}
\caption{Average photon number $N_o(t_p)$ hitting the odd detector (D)
  for $N_{in}=10^9$ incoming photons, as a function of the delay time
  $t_p$ in microsecond at which the measurement takes place.  In the
  inset the initial oscillations are magnified. Parameters are given
  in Table~\ref{t:parameters} (Rb atom).
The coupling to the environment is taken in the pure decoherence form
of Eq.(\ref{eq:dephasing}). A thermalizing environment
[Eq.(\ref{eq:thermalization})] would not lead to
the superimposed oscillations, but to an otherwise similar behaviour.
}
\label{signal}
\end{center}
\end{figure}

The signal in the odd photon detector given by Eqs.\
(\ref{eq:Nodd-result}) and (\ref{eq:result-sudden-1}) is plotted in
Figure~\ref{signal}, using the spatial decoherence
model~(\ref{eq:dephasing}) and the parameters given in
Table~\ref{t:parameters}. The signal shows an overall increase because
the environment heats the system and broadens its position
distribution. The important feature of this plot are the superimposed
oscillations.  They stem from the breathing motion of the wave packet
and are a telltale sign of spatial decoherence that affects position
and momentum in a non-equivalent manner.  Indeed, Eq.~(\ref{eq:ddtp2})
shows that decoherence only increases the momentum width which leads
to a squeezed phase space distribution.  The dynamics in the harmonic
potential makes this elongated distribution rotate at the frequency
$\Omega$ so that its projection onto the position or momentum axis
oscillates in width at an angular frequency $2\Omega$ (as expected
from Eqs.\ \ref{eq:ddt_cov}).

\subsection{Signal visibility}
\label{s:s/n}

Next, we discuss ways to extract the oscillations at $2\Omega$ of
the odd detector signal that are characteristic for spatial
decoherence.

Let us consider probing times $t_p$ in the range where the system
is in the Lamb-Dicke limit.  That is, in Fig.\ \ref{signal}, we
are well before the saturated regime, i.e. $t_p< 100 \mu$s.  The
number of odd photons is then given by Eq.\
(\ref{eq:sin2-LambDicke}) which, combined with Eq.\
(\ref{eq:solution-x2}) yields the peak-to-peak amplitude of the
breathing oscillations
\begin{equation}
    \left. N_{o} \right|_{\rm osc}
    = 8 N_{in} R (k \sigma_{0})^2 \frac{ \Gamma }{ \Omega }
    ,
    \label{eq:signal-3}
\end{equation}
where $(k \sigma_{0})^2 = \hbar k^2 / 2 m \Omega$ is the so-called
Lamb-Dicke parameter.  It is interesting to note that the oscillation
amplitude~(\ref{eq:signal-3}) depends neither on the initial system
state nor on the delay time $t_{p}$ for the probe
pulse~\cite{expectFriction}. However, the last three factors, $R$, $k
\sigma_{0}$, and ${\Gamma }/{\Omega}$ are all less than unity, and a
significant photon number can only be obtained with a bright probe
pulse, $N_{in} \gg 1$. In the example given above, $N_{in} \sim
10^{9}$ for a 20\,ns pulse ($10^{-10}$ J per pulse), and this large
number compensates for the small fraction of `useful photons' $R
\approx (g/\kappa)^2 \sim 10^{-8}$.  The signal can be improved a lot
with a larger coupling strength $g$. For the optimal value $g =
\kappa$, the ratio $R$ takes its maximum value $1/4$.  The signal
amplitude can also be increased with a shorter probe wavelength as
long as one remains in the Lamb-Dicke limit.

The contrast $C_{\rm osc}$ of the breathing oscillations is defined by
dividing the oscillation amplitude~(\ref{eq:signal-3}) to the smoothed
background signal. Since that background increases with $t_{p}$ (in
the Lamb-Dicke regime), the contrast is time-dependent. In terms of
the initial system temperature $T_{0}$,
\begin{equation}
    C_{\rm osc}( t_{p} ) =
    \frac{ 2 \Gamma / \Omega }{
    \coth( \hbar \Omega / 2 k_{B} T_{0} ) +
    \Gamma t_{p} }
    \;,
    \label{eq:signal-contrast-1}
\end{equation}
where the dependences on photon number, detection probability and
probe wavelength have canceled.  The initial contrast is maximized
for low temperatures $k_{B} T_{0} \le \hbar \Omega$ when the
system is cooled close to the ground state.  Even then it is
limited by the small ratio $\Gamma / \Omega$, see
Eq.~(\ref{eq:condition-master-eqn-2}).  For higher temperatures,
the contrast decreases like $\hbar\Gamma / k_{B} T_{0} \ll 1$.
The contrast reduction with increasing $t_{p}$ happens on the time
scale $\coth( \hbar \Omega / 2 k_{B} T_{0} ) / \Gamma$, where
decoherence has approximately doubled the position variance
compared to its initial value.  This time scale is of order
$1/\Gamma$ if the initial temperature is low. Otherwise, when the
initial position distribution is already quite broad, it is of
order $k_{B} T_{0} / ( \hbar\Omega \Gamma)$, so the decoherence
needs longer to double the initial width.

To get a realistic estimate for the visibility of the oscillations,
however, one has to take into account the fluctuations of the detector
signal. The variance $\Delta N_{o}^2$ is given in
Eq.\ (\ref{eq:Nodd-variance-2}) for the Lamb-Dicke limit. It shows a
super-Poissonian scaling with $N_{in}^2$ to leading order so that we
get a signal to noise ratio
\begin{equation}
    S/N( t_{p} ) = \frac{ N_{o}|_{\rm osc} }{ \Delta N_{o} } =
    \frac{ C_{\rm osc}( t_{p} ) }{ (2 + 3 / N_{in})^{1/2} }
    \;,
    \label{eq:S/N-1}
\end{equation}
which is much smaller than unity. In order to resolve the
oscillations, the S/N ratio has to be enhanced by repeating the
experiment a number of times
\begin{equation}
  \label{eq:numb_of_exp}
   n_{\rm ex} \approx 10 / [S/N(t_{p})]^2  \;,
\end{equation}
which ensures that the measured oscillation amplitude at an antinode
exceeds the background noise level by three standard deviations. For
short enough times this condition yields a number of $n_{\rm ex} \sim
(\Omega / \Gamma )^2 \gg 1 $ recordings for a given time $t_p$.

Alternatively,  the  oscillations can  be  extracted  from the  signal
frequency spectrum in terms of the  peak they give at $2\Omega$. If we
assume that the  increase of the background is  negligible on the time
scale $2\pi  / \Omega$, the signal fluctuations  give an approximately
white noise. In terms of the  weight of the peak at $2\Omega$, we find
an improvement  by a factor  $(T/\Delta t)^{1/2}$ with respect  to the
signal  to  noise ratio~(\ref{eq:S/N-1}),  where  $T$  is the  maximum
probing time  in the  data set  and $\Delta t  \approx \tau$  the time
resolution.  This  shows that the sudden approximation  (a small value
of $\tau$)  is actually required to  extract the signal.   We can take
into account the slow increase of the background in an approximate way
by writing the spectral signal to noise ratio as
\begin{eqnarray}
    S/N( 2\Omega ) &\approx& ( T / \tau )^{1/2} S/N( t_{p} = T / 2 )
    \label{eq:S/N-2}
    \\
    &=&
    \frac{ ( T / \tau )^{1/2} }{ (2 + 3 / N_{in})^{1/2} }
    \frac{ 2 \Gamma / \Omega }{
    \coth( \hbar \Omega / 2 k_{B} T_{0} ) +
    \Gamma T / 2 }
    \;.
    \nonumber
\end{eqnarray}
The signal to noise ratio shows a maximum for a specific observation
time
\begin{equation}
  \label{eq:Topt}
  T_{\rm opt} = \frac{\coth( \hbar \Omega / 2 k_{B} T_{0} )}{\Gamma}
\end{equation}
which is $T_{\rm opt} \sim 1 / \Gamma$ for low initial
temperatures. It is interesting that the system should be monitored up
to the time where the contrast~(\ref{eq:signal-contrast-1}) of the
oscillations decreases. As a characteristics of the efficiency of the
detection process, we obtain the maximum S/N
\begin{equation}
    \max S/N( 2\Omega ) \approx
    \left( \frac{ \Gamma  \tanh( \hbar \Omega / 2 k_{B} T_{0} ) }{
    \Omega^2 \tau } \right)^{1/2}
    \;.
    \label{eq:S/N-opt}
\end{equation}
The spectral signal to noise ratio can be significant, i.e.\ we get $\max S/N(
2\Omega ) > 3$ for the system parameters of Table \ref{t:parameters},
and it thus provides an evidence for the oscillations. Note, however,
that this signal is obtained from the records for the whole time
series from 0 to $T_{\rm opt}$. That is, a measurement repetition
$n_{\rm ex} \approx T_{\rm opt} / \tau \approx 1/(\Gamma \tau)$ is
necessary. On taking into account that a time resolution $\tau < 0.1
\, \Gamma / \Omega^2$ is required for getting such a significant S/N,
we arrive at a necessary number of measurements $n_{\rm ex} \approx
(\Omega / \Gamma )^2 \gg 1$. This condition is similar to the one we
obtained for the resolution of the oscillations at a single point
$t_p$. This result also justifies the use of a controlled decoherence
source to enhance the decoherence rate $\Gamma$. For example, with the
parameters given in Table \ref{t:parameters}, about $10^4$ measurements
are required for evolutions in the range of 0 to $100\,\mu$s to get
the signal to noise ratio larger than 3.

\subsection{Example II: Nanoparticle}

Finally, we come back to the main motivation of this work and
consider the decoherence of massive systems.  As examples of such
a nanoparticle we think of recently realized nano
electromechanical oscillators (NEMs) \cite{huang03} or beads,
which may be put in an harmonic confining potential by holding
their edges or by suspending them in midair by electric (e.g. Paul
trap in case they are charged), magnetic (in case they posses a
magnetic moment which maintains a meta stable state), or optical
interaction (optical tweezers for small beads).

The crucial parameter is the initial temperature which makes a
difference with respect to the case of single atoms. While atoms can
be routinely cooled down to the ground state of oscillation in a trap
by various optical methods, there is no such efficient cooling scheme
for arbitrary massive objects. We can safely assume that nanoparticles
can be cooled down to below 1K or even, in the near future, to the mK
range, using e.g.\ special cryogenics or opto-mechanical
feedback \cite{feedback} (for  experiments, see Cohadon et al.
in Ref.\cite{optomech}).
With experimentally feasible trapping frequencies in the
range of MHz, such temperatures are still large, $\hbar\Omega \ll k_B
T_0$, and the initial vibrational state of the system is highly
excited.

On the other hand, for large enough masses, such a temperature range
is sufficiently low to reach the Lamb-Dicke regime,
\begin{equation}
  \label{eq:LambDicke_massive}
  k^2 \langle x^2 \rangle_0 = (k \sigma_{0})^2 \,
  \frac{2 k_B T_0}{\hbar \Omega} \ll 1\, ,
\end{equation}
where again $(k \sigma_{0})^2 = \hbar k^2 / 2 m \Omega$ is the
Lamb-Dicke parameter (see Eq.(\ref{eq:signal-3})).
According to Eq.\ (\ref{eq:result-sudden-1}), this is the basic
necessary condition for the proposed detection scheme to work.  We can
then use the general result of Eq.\ (\ref{eq:S/N-opt}) and expand the
tanh function to lowest order:
\begin{equation}
  \label{eq:S/N_massive}
  \max S/N( 2\Omega ) \approx \left( \frac{\hbar \Gamma}{2 k_B T_0 \, \Omega \tau}
  \right)^{1/2}  \; ,
\end{equation}
and similarly from Eq.\ (\ref{eq:Topt}):
\begin{equation}
  \label{eq:Topt_massive}
  T_{\rm opt} \approx \frac{2 k_{B} T_{0}}{\hbar\Omega\Gamma} \, .
\end{equation}
The requirements for the experiment can be quantitatively estimated on
the basis of these equations, regardless of the specific
implementation of the scheme.

One important quantity is the number of measurements needed to get a
statistically significant signal of the oscillations in the odd photon
detector. First, a number of $T_{\rm opt}/\tau$ measurements has to be
carried out to monitor the decoherence evolution from $t_p=0$ to
$t_p=T_{\rm opt}$. Second, this full evolution has to be recorded a
number of about $10 / (\max S/N)^2$ times to reach, by spectral
analysis, three standard deviation from the noise level. The total
number of measurements is then
\begin{equation}
n_{\rm ex} \approx 10/(\max S/N)^2 \times T_{\rm opt}/\tau
\sim 10
\, (2 k_B T_0/\hbar\Gamma)^2\, ,
\label{eq:nex-estimate-np}
\end{equation}
which heavily depends on the initial temperature. It follows that the
decoherence rate should be artificially enhanced to the maximum level
allowed by the condition (\ref{eq:condition-master-eqn-1}), i.e.\
$\Gamma = 0.1\,\Omega$.

\begin{table}
    \begin{tabular}{|l|c|}
        \hline
        mass $m$ & $ < 10^{-15}$ {\rm kg}
        \\
        vibration frequency $\Omega / 2\pi$ & 1\,{\rm MHz}
        \\
        pulse length $\tau$ & 30\,{\rm ns}
        \\
        decoherence rate $\Gamma/2\pi$ & $100\,{\rm kHz}$
        \\
        initial temperature $T_0$ & 1 mK
        \\
        \hline
        max S/N & 0.1
        \\
        $T_{\rm opt}$ & 100 $\mu$s
        \\
        \hline
        $n_{\rm ex}$ & $10^6$
        \\
        \hline
        \end{tabular}
        \caption{First block: parameters for a reference nanoparticle
        fulfilling the validity conditions of the calculation.  Second
        block: calculated maximum of the `signal to noise' ratio of the
        spectral analysis and the maximum recorded evolution time.
        Third block: the estimated number of experimental runs in order
        to achieve three standard deviations and the verification of
        the decoherence signal.}
        \label{t:nanoparameters}
\end{table}

Estimates for a reference nanoparticle are assembled in Table
\ref{t:nanoparameters}. For a trapping frequency $\Omega = 2\pi\,
1$ MHz, the decoherence rate can be as large as $\Gamma = 2\pi\,
100$ kHz.  In order to maximize the signal to noise ratio, the
pulse length can be as short as $\kappa\tau \approx 10$. For a
cavity linewidth $\kappa = 2\pi \, 50$ MHz, as the one in the Rb
atom example, an initial temperature $T_0=1$ mK yields the ratio
$\max S/N( 2\Omega ) \approx 0.1$, and we get a number $n_{\rm ex}
\approx 10^6$ where one measurement takes a time of about $T_{\rm
opt}/2 \approx 100 \mu$s (not including the preparation time).

In many possible physical realizations, the coupling to the
nanoparticle is independent of its mass, e.g. the spring constant
of the trap, $K = m \Omega^2$, and the parameter $D$ of the
decoherence functional~(\ref{eq:dephasing}). It is very
interesting that $\Gamma/\Omega = D/K$ and, hence, the maximum
signal to noise ratio in Eq.\ (\ref{eq:S/N_massive}) is
independent of the mass of the particle. This proves that the
scheme can be extended to monitor the decoherence of massive
particles. The mass, in fact, can be scaled up to a limit $m_{\rm
cr}$ without degrading the signal.  This limit is related to our
assumption that the particle remains smaller than the wavelength
so that it couples to the light field via its polarizability.  The
upper mass limit is $m_{\rm cr} \equiv \varrho \lambda^3 \sim
10^{-15}\,{\rm kg}$, i.e.\ about $10^{10}$ Rb atoms, for a typical
material density and visible light.  The optimum time period to
reach the maximum spectral visibility, on the other hand, depends
on the mass $T_{\rm opt} \sim m \, 2k_{B}T_{0}/D$ and makes the
necessary conditions of observing decoherence less demanding for
smaller masses.

\section{Conclusion}

To conclude, we have addressed the problem of observing the
decoherence process in massive objects. Experimentally quantifying the
process for large masses and different environments is of paramount
importance for the accurate theoretical modeling of this subtle
transition from quantum to classical.

Specifically, we have shown that it should be possible to study
the decoherence of systems that are trapped.  In contrast to
conventional spatial dephasing experiments, in which the observed
object is freely propagating, the presented scheme will allow for
more isolation, and hence for a better control over the coupling
to the environment. This is made possible by making use of a
photon probe that scans the object for changes in its wave
function.  Furthermore, the suggested experimental scheme,
bypasses the need to create a clearly separated spatial
superposition. This need presents an ever growing technical
difficulty for large masses in terms of the actual preparation
procedure and in terms of the rate of decoherence. A ``pure"
decoherence signal is apparent even when no superposition was
created initially.  The only requirement in the present scheme is
to prepare the system in an equilibrium state of an harmonic
potential. This is a further advantage over the ground state
cooling required by numerous other schemes.

These features will greatly enhance the feasibility of a
decoherence experiment with a scalable object mass. In addition,
the interaction with optical cavity modes can be used to tailor
the object's environment and to induce ``decoherence on demand"
\cite{MaiaNeto00}.

In this paper, we have limited ourselves to objects smaller than
the wavelength that weakly scatter light. Future work will address
moving mirrors that lead to a stronger optical signal. The fact
that our scheme does not require strong reflectivity will allow
the use of very light double sided mirrors (foils) with high
transmittance. Preliminary results show that a similar Hamiltonian
accounts for the coupling to the cavity modes so that most of the
present analysis can be carried over, but with more favorable
parameters.

Finally, we have focused on environmental decoherence. However,
if the system may be isolated well enough so that the
coupling to the environment does not mask other weaker processes
of localization, then one may perhaps be able
to study also other proposed models \cite{GRW,penrose}.
For estimates
of the parameters required for opto-mechanical tests,
see Bose et al. and Marshall et al. in Ref.\cite{knight}.

\acknowledgments

C.\ H.\ and M.\ N.\ thank J.\ Eisert (Potsdam) for stimulating
discussions. P.\ D.\ acknowledges the support by the National
Scientific Fund of Hungary (Contracts No. T043079 and T034484) and
that of the Bolyai Fellowship Programme of the Hungarian Academy of
Sciences. R.\ F.\ would like to thank for their support the German
Federal Ministry of Education and Research (BMBF), the Israel Science
Foundation, and the European Union Research Training Network
(MRTN-CT-2003-505032).

\appendix

\section{Interaction Hamiltonian}

We describe the interaction between the electromagnetic field and the
object by $\frac12 \alpha {E}^2$, where $\alpha$ is the polarizability
and $E$ the (linearly polarized) electric field.  In second
quantization, keeping only the even and odd cavity modes, the field
can be written
\begin{equation}
E = \sqrt{ \frac{\hbar\omega_{c}}{\epsilon_0V} }
\{ a_e \cos(kx) +a_o \sin (kx) + h.c.
\} ,
\end{equation}
where $\omega_{c}$ is the (common) mode frequency, the wavenumber $k =
\omega_{c} / c$, $V$ is the cavity mode volume.  In the rotating wave
approximation and adopting normal order, we hence find for the
interaction
\begin{eqnarray}
H_{int} &=& \frac{\alpha \hbar \omega_{c} }{2 \epsilon_{0} V }
\left\{ a_e^\dagger a_e + a_o^\dagger
a_o
+ \left( a_e^\dagger a_e - a_o^\dagger a_o \right) \cos (2kx)
+ \right.
\nonumber\\
&& \left. {}
+ \left( a_e^\dagger a_o +a_o^\dagger a_e \right) \sin (2kx) \right\}
.
\label{eq:atom-interaction}
\end{eqnarray}
We see here that the scattering from even into odd modes is
accompanied by an anti-symmetric excitation of the object.
If the photon leaves the cavity in the same mode, however, the
object state is changed by the symmetric function $\cos(2kx)$.
For the interaction Hamiltonian used in the main text, we
have left out the part $a_e^\dagger a_e + a_o^\dagger a_o$
involving the total photon number. Since this is a conserved
quantity, it only contributes a global phase factor to the
system+field wave function.

Writing $\hbar g$ for the prefactor in Eq.\ 
(\ref{eq:atom-interaction}), we get the coupling constant $g$.  With
an atom as system, we adopt a two-level model for the AC
polarizability and cast the coupling in the form
\begin{equation}
    g = \frac{ 3 \omega_{c} }{ 8\pi \omega_{A} } \nu
    \frac{ \gamma_{A} }{ \omega_{A} - \omega_{c} }
    \frac{ \lambda_{A}^2 }{ A }
    \label{eq:estimate-g}
\end{equation}
where $\omega_{A}$ is the atomic resonance frequency andq
$1/\gamma_{A}$ the corresponding radiative lifetime. $A$ is the cross
section of the cavity modes and $\nu = c / L$ the free spectral
range. Eq.\ (\ref{eq:estimate-g}) assumes an excitation not too far off
resonance, $|\omega_{A} - \omega_{c}| \ll \omega_{A}$.  Typical
parameters are $\nu \sim 10^9/{\rm s}$, $A / \lambda_{A}^2 \sim
10^3$, $|\omega_{A} - \omega_{c}| / \gamma_{A} \sim 10$, and give $g
\sim 10^4/{\rm s}$. This can be increased further with tighter
focussing, smaller cavity lengths and working closer to resonance. A
far off-resonant excitation with a shorter wavelength would improve
the resolution of the position measurement. This is likely to be
over-compensated by the smaller polarizability, since for $\omega$
well above a resonance transition, $\alpha \propto 1/\omega^2 \to 0$.

If the system is a nanoparticle with size smaller than the wavelength,
the polarizability can be written in the Clausius-Mossotti form
\begin{equation}
    \alpha( \omega ) = 4\pi\varepsilon_{0} a^3 \frac{ \varepsilon(
    \omega ) - 1 }{ \varepsilon(
    \omega ) + 2 },
    \label{eq:nano-part-alpha}
\end{equation}
where where $\varepsilon( \omega )$ is the particle's permittivity.
The factor $\mu = ( \varepsilon( \omega ) - 1 )/( \varepsilon( \omega
) + 2 )$ shows a resonance for example in metallic particles
(collective plasma oscillation). A value $\mu \sim 5$ seems realistic
while avoiding too large absorption losses.  This gives a coupling
\begin{equation}
    g = (2\pi)^2 \mu \,\nu \frac{ \lambda^2 }{ A }
    \frac{ m }{ m_{\rm cr} },
    \label{eq:coupling-nano-part}
\end{equation}
where the `critical mass' is defined in terms of the mass density
$\varrho$ as $m_{\rm cr} = \varrho \lambda^3$. Typical numbers
($\varrho = 1\,{\rm g / cm}^3$, $\lambda = 1\,\mu{\rm m}$) give
$m_{\rm cr} \sim 10^{-15}\,{\rm kg}$.  With the same numbers for the
cavity mode as before, we get the fairly large value $g \approx 
\nu (m / m_{\rm cr})$.


\begin{thebibliography}{10}
\bibitem{expectFriction}
We expect, however,
the oscillations to be damped at times where friction,
that has been neglected in the
pure decoherence functional~(\ref{eq:dephasing}),
sets in.
\end{thebibliography}
\end{document}